# Epithelial-substrate coupling strength regulates the landscape of the traction in cohesive monolayers: a parametric study and a revisit to "size effect"

Tiankai Zhao*, Hongyan Yuan

Shenzhen Key Laboratory of Soft Mechanics & Smart Manufacturing, Department of Mechanics and Aerospace Engineering, Southern University of Science and Technology, Shenzhen, 518055, China

*Corresponding Author:   zhaotk@sustech.edu.cn

## Abstract

Epithelial cells can assemble into cohesive colonies and collectively interact with substrates by generating extracellular forces through focal adhesions. Recently, a molecularly based thermodynamic model, which integrates both the monolayer elasticity and force-mediated focal adhesion formation, has been developed to elucidate the regulation of the cellular force landscape induced by the active epithelial-substrate coupling. However, how epithelial-substrate coupling strength mediate the landscapes of the traction, the cellular displacement, and the focal adhesion distribution in a cohesive monolayer remains unexamined in details. In this work, we follow the procedures by the previous work to re-formulate the free energy of the epithelial-substrate system and obtain the thermodynamic steady-state equations. We then derive a simplified form of the complete equation system, and solve it both semi-analytically and numerically. We find that the parameter which characterizes the epithelial-substrate coupling strength can significantly affect the landscapes of the traction the cellular displacement, and the focal adhesion distribution. We also revisit the "size effect" addressed by previous works and demonstrate that such effect is the natural outcome of a strong epithelial-substrate coupling without introducing any extra factors. For epithelial-substrate coupling which is not strong enough, the currently observed "size effect" does not hold. A scaling law that determines whether the previously observed "size effect" holds is proposed based on our model.

1. Introduction

Living cells have the ability to generate, transmit, and sustain mechanical forces.(1) In a multicellular system, individual cells generate intracellular contraction by actomyosin motors. The contraction is transmitted by stress fibers either to neighboring cells through E-cadherin-based cell-cell junctions, generating intercellular tension; or to the surrounding matrices through focal adhesions, generating extracellular traction.(2) The extracellular tractions then act back onto the multicellular tissues and help cells maintain their internal stress state, called stress homeostasis.(3) To characterize such cell-substrate coupling has drawn great interests in biophysical and biomechanical communities.(4) A well-known experimental technique to achieve this goal is the so-called Traction-force-microscopy (TFM), which tracks the displacement in the substrate by implanting fluorescent beads and solves an inverse elastic problem based on the measured displacement field.(5–7) This method, although being effective in measuring the extracellular tractions, is quite time-consuming, has limited sensitivity, and cannot be applied to rigid substrate. To overcome these restrictions and gain scientific insights on how cells actively interact with their surroundings, biophysicist and mathematicians have developed quite a lot of theoretical and computational models to study the active coupling between adherent cells and substrates. On the molecular level, the kinetics of focal adhesions has been widely studied by a large number of models that incorporated the mechanosensitivity of such structures(8–11). With experimental observations and mechanics modeling, Yuan et al.(12) revealed that focal adhesion distribution can affect the spatial organizations of stress fibers, in addition to the cell shape. On the single cell levels, Deshpande et al.(13–15) developed a series of bio-mechanical models for coupling cell contractility, stress fiber assembly, and focal adhesion formation. The models successfully predicted a large number of experimental observations, such as the sensitivity of the contractility response to the substrate stiffness. Later, Oakes et al.(16) studied the traction profile within isolated adherent cells with different geometries by a minimal model. He et al.(17) then studied this model and derived the semi-analytical solutions to the model with respect to different substrate thickness. Shenoy et al.(18) developed a chemo-mechanical free-energy-based model that explains the durotaxis and extracellular-stiffness-dependent contractility. Recently, Yang et al.(19) have adopted Deshpande's model to study the fibroblasts adhering on substrate with different stiffness and thickness. On the multicellular tissue level, Mertz et al.(20) obtained a scaling law of traction forces with the size of cohesive cell colonies. Banerjee and Marchetti(21) studied the contractile stress in cohesive cell monolayers on substrate with different thickness. Mertz et al.(22) studied how cadherin-based cell-cell junctions could affect the traction distributions in the epithelial monolayers. Later, Zhao et al.(23) developed a molecular-based thermodynamic model, integrating monolayer and substrate elasticity, and force-mediated focal adhesion to elucidate the active coupling. This model was used to predict the traction distribution in irregular-shaped cohesive colonies with different sizes seeded on substrates with varying stiffness.

Recently, Zhang et al.(24) have systematically studied the traction profile of cohesive monolayers and built a correlation between the average traction within the 50 μm-wide boundary layer and metastasis. The HCT-8 cells were seeded on 20.7 kPa PAA gels and formed into nearly-rounded cohesive colonies with different radiuses $R$. They measured the tractions generated by these colonies and discovered the so-called "size-effect". Such effect says: on the substrate with the same stiffness, smaller-sized cohesive colonies possess larger average traction within boundary layers than the bigger-sized colonies. This effect was later explained by the possible existence of the large line tension on the cell boundary.(25) In the hypothesis, the line tension $\Gamma$, which was assumed as a material constant(25), could generate an effective inward pressure proportional to the boundary curvature: $p \sim \Gamma \kappa / h_c$ with $\kappa$ and $h_c$ being the boundary curvature and monolayer thickness. As the size of colonies $R$ decreases, the boundary curvature $\kappa \sim 1/R$ increases. The inward effective pressure $p$ increases dramatically, asking for more traction to counterbalance, thus leading to an increase in the average traction. The work has predicted that a line tension with a value in the order of $\sim 10^{-7}$ J/m could lead to the size effect observed in the experiment. Yet whether the line tension could possess such a value still requires experimental measurements. In the present work, we will show that the size effect observed in previous experiments can be explained without introducing the line tension.

In this paper, we will focus on how epithelial-substrate coupling strength affects the landscapes of cellular forces, displacements, and focal adhesion distributions in a cohesive epithelial monolayer. To accomplish this, a detailed description of the settings of our molecular-based continuum-level model is firstly presented by following the work of Zhao et al.(23); a free energy functional for a cohesive epithelial monolayer colony adhering on substrate is then formulated and the thermodynamic steady-state is obtained thereafter. The parameter that characterizes the strength of epithelial-substrate coupling is identified. We then simplify the steady-state equation system based on a few assumptions and solve it semi-analytically for the case that the epithelial-substrate coupling is quite weak. This is followed by a detailed parametric study on how the epithelial-substrate coupling strength could affect the profiles of the traction, the displacement, and the focal adhesion distribution within the colony. Finally, we apply the model to study the average traction within the boundary layers of the nearly-circular colonies that possess different sizes. We study the "size effect" raised by previous works and present a brief discussion on our predictions and previous experimental observations. A scaling law that results in the "size effect" is proposed at the end of the article.

**2. The thermodynamic steady-state equations for adherent cohesive epithelial monolayer**

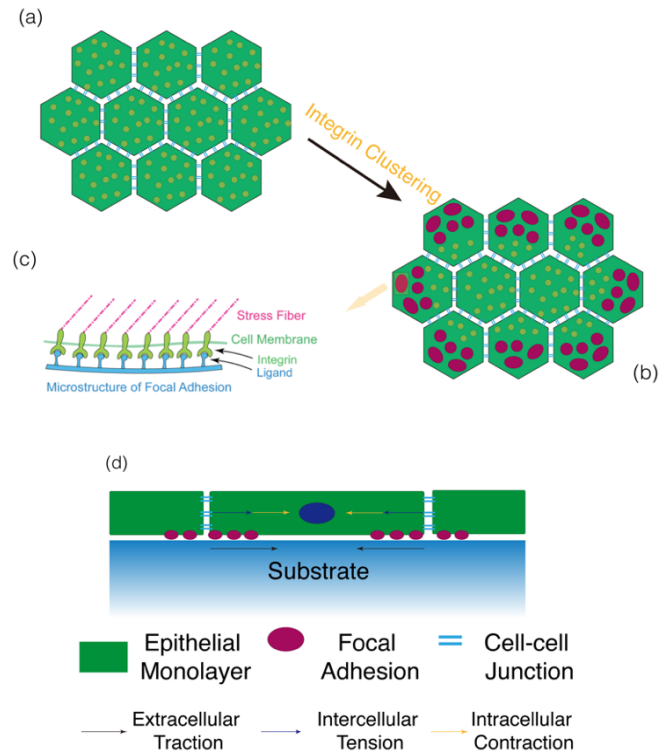

**Fig. 1** Illustration of a cohesive epithelial monolayer colony adhering on substrate. (a) The integrins, represented by small orange dots, are initially randomly distributed on cell membrane; (b) Once upon the cells adhere to the substrate, large focal adhesion points, represented by large red dots, will form through clustering; (c) Molecular structures of focal adhesions; (d) A side view of how the epithelial monolayer generate and transmit cellular forces.

In this section, we follow the procedures in previous works to formulate the thermodynamic steady-state equations for cohesive epithelia monolayers adhering onto a substrate. To a minimal setting, we consider an epithelial monolayer colony with a total of $N$ cells seeded on substrates coated with fibronectin to which integrins bind and form focal adhesion points (see Fig. 1). The extracellular traction and intercellular tension originated from actomyosin contraction displace the monolayer. Adopting a continuum view within the framework of elasticity, we denote the displacement fields of the monolayer colony by $\mathbf{u}$, and the respective conjugate Cauchy stress fields by $\boldsymbol{\sigma}$. Focal adhesion points, the clusters of ligand-integrin complexes, transmit forces between the monolayer and the substrate. In order to capture the molecular basis of focal adhesion formation, the integrin receptors on the *i*-th cell membrane are categorized into two distinct phases: a freely diffusive phase with a normalized density of $\phi_{i,f}$, and a bound phase to the ligands on the substrate with a normalized density of $\phi_{i,b}$, which can be used to characterize the focal adhesion density. The two phases, together, satisfy the local inequality: $\phi_{i,f} + \phi_{i,b} < 1$. Conservation of the receptors expressed on the membrane within each cell requires $\iint (\phi_{i,f} + \phi_{i,b}) d\Omega_i = \phi_{i,0} \Omega_i$, with $\phi_{i,0}$ being a constant for the *i*-th cell that covers an area of $\Omega_i$. Among these phases, only the bound phase contributes to the focal adhesion points and transmits the force through clustering. Within the focal adhesion points, each receptor-ligand pair is modeled as a Hookean spring with an effective spring constant $k_e$, where $k_e$ is the effective spring constant, standing for a joint effect brought by both the integrin-ligand complex itself and the substrate. The stretching force in the pair is $\mathbf{F} = k_e \mathbf{u}$, where the stretch $\Delta \mathbf{u}$ in the spring is approximated as $\Delta \mathbf{u} \approx \mathbf{u}$. Accordingly, the traction force per unit area generated by the receptor-ligand pair and transmitted to the substrate is $\mathbf{T} = n\phi_b \mathbf{F}$, where $n$ is the lattice site per unit area on the cell membrane and $\phi_b = \{\phi_{i,b}\}$ lumps the spatially organized focal adhesion points of all the constituent cells.

With the above settings, the free energy functional of the cohesive monolayer colony takes the following form:

$$f = \sum_i^N W_i(\phi_{i,f}; \phi_{i,b}; \mathbf{u}) + \frac{h_c}{2} \iint \boldsymbol{\sigma} : \boldsymbol{\varepsilon} \mathrm{d}\Omega, \tag{1}$$

where the first term is the summation of the free energies associated with the receptors of all the cells, and the second term is the elastic strain energy within the entire colony, which is modeled as a two-dimensional elastic sheet with a thickness $h_c$ and occupying an area $\Omega$. Here, we note for the cohesive epithelial monolayer we model, the cell-cell junction is much stiffer than the cells themselves and thus can hardly deformed by the intracellular contraction. Hence, the displacement field $\mathbf{u}$ is continuous over the whole cell colony, and the energies possessed by the cell-cell junctions are negligible. In Eq. (1), the first term is written as:

$$W_i(\phi_{i,f}; \phi_{i,b}; \mathbf{u}) = \iint n(\mu_{i,f}^0 \phi_{i,f} + \mu_{i,b}^0 \phi_{i,b}) \mathrm{d}\Omega_i + \iint n k_B T [\phi_{i,f} \ln \phi_{i,f} + \ln \phi_{i,b} + (1 - \phi_{i,f} - \phi_{i,b}) \ln (1 - \phi_{i,f} - \phi_{i,b})] \mathrm{d}\Omega_i + \frac{1}{2} \iint \left[ \lambda_{i,f} (\nabla \phi_{i,f})^2 + \lambda_{i,b} (\nabla \phi_{i,b})^2 \right] \mathrm{d}\Omega_i + \frac{1}{2} \iint n \phi_{i,b} k_e |\mathbf{u}|^2 \mathrm{d}\Omega_i - \iint n \phi_{i,b} \gamma \mathrm{d}\Omega_i. \tag{2}$$

The free energy of the receptors on the cell membrane consists of five components. The first integral in Eq. (2) stands for the chemical energy in the reference configuration, where $\mu_{i,f}^0$ and $\mu_{i,b}^0$ are the referential chemical potential for $\phi_{i,f}$ and $\phi_{i,b}$, respectively; $n$ is the number of lattice sites per unit area on the cell membrane. The second integral denotes for the entropic energy of the two-phases mixtures, where $k_B$ is the Boltzmann constant, and $T$ is the temperature. The third integral represents for the gradient energy at the interfaces of the two phases, where $\lambda_{i,f}$ and $\lambda_{i,b}$ are the gradient energy coefficient with respect to the two phases. The fourth integral is the stretch energy stored in the ligand-receptor complexes that are modeled as Hookean springs. The last integral is the energy release upon the formation of each ligand-receptor complex, which serves as a stabilization term with respect to the free energy. The larger the value of $\gamma$ possesses, the more focal adhesion assemblies are in favored. The energy release is assumed to have the form $\gamma = \mathbf{F} \cdot \mathbf{u}$: an analogy to the $-PV$ term in the thermodynamics of gases.(15)

Minimizing the free energy functional with respect to its independent variables $\phi_{i,f}$ and $\phi_{i,b}$ under the constraint $\iint (\phi_{i,f} + \phi_{i,b}) d\Omega_i = \phi_{i,0} \Omega_i$ yields the chemical steady-state in the $i$-th cell:

$$-\lambda_{i,f} \Delta \phi_{i,f} + nk_B T \ln \frac{\phi_{i,f}}{1 - \phi_{i,f} - \phi_{i,b}} + n\mu_{i,f}^0 + L_i = 0, \tag{3}$$

$$-\lambda_{i,b} \Delta \phi_{i,b} + nk_B T \ln \frac{\phi_{i,b}}{1 - \phi_{i,f} - \phi_{i,b}} - \frac{1}{2} nk_e |\mathbf{u}|^2 + n\mu_{i,b}^0 + L_i = 0, \tag{4}$$

where $L_i$ is the Lagrangian multiplier. The boundary conditions for Eq. (3) and (4) are set to be:

$$\nabla \phi_{i,f} \cdot \mathbf{n}_i = 0, \tag{5}$$

$$\nabla \phi_{i,b} \cdot \mathbf{n}_i = 0, \tag{6}$$

where $\mathbf{n}_i$ is the outer unit normal of the $i$-th cell boundary. By subtracting Eq. (4) from Eq. (3), one can obtain the partition of receptors in each cell:

$$\lambda_{i,b} \Delta \phi_{i,b} - \lambda_{i,f} \Delta \phi_{i,f} - nk_B T \ln \frac{\phi_{i,b}}{\phi_{i,f}} + \frac{1}{2} nk_e |\mathbf{u}|^2 + n(\mu_{i,f}^0 - \mu_{i,b}^0) = 0. \tag{7}$$

Minimizing the free energy functional with respect to the displacement field $\mathbf{u}$ gives rise to the mechanical equilibrium condition of the whole cell sheet:

$$\nabla \cdot \boldsymbol{\sigma} - nk_e \phi_b \mathbf{u}/h_c = \mathbf{0}. \tag{8}$$

By liking into Eq. (7) and (8), one can see that $k_e$ characterizes the coupling between the monolayer and the substrate. The boundary condition for the mechanical equilibrium is written as:

$$\sigma \mathbf{n} = \mathbf{0}, \qquad (9)$$

where $\mathbf{n}$ is the outer unit normal the of the boundary of the monolayer colony. We model the epithelial colony as a two-dimensional isotropic elastic sheet, with Young's modulus $E$ and Poisson's ratio $\nu$. The intracellular contraction is modeled by an active pressure $\sigma_A$ within the epithelium, resembling a tension induced by thermal cooling. The constitutive relation of the monolayer is set as:

$$\sigma_{ij} = \frac{E}{1-\nu^2}\left[\nu \delta_{ij} u_{k,k} + \frac{1-\nu}{2}(u_{i,j} + u_{j,i})\right] + \sigma_A \delta_{ij}, \qquad (10)$$

where a small deformation case is assumed in Eq. (10).

Retrieving the cell-cell boundaries and solving the equation system within each cell could be quite expensive for large colonies. Hence, to lower down the complexity and make the equation system more mathematically tractable, we can simplify it by making a few approximations. The first and the most important approximation is that: for a cohesive monolayer colony whose radius ($R \sim 10^2$ μm) is much larger than the size of each cell ($r_c \sim 10^0 - 10^1$ μm), the area occupied by the $i$-th cell is much smaller than the area of the entire monolayer: $\Omega_i \ll \Omega$; hence the conservation of the receptors on the membrane within each cell can be approximated as:

$$\iint (\phi_{i,f} + \phi_{i,b}) d\Omega_i \approx (\phi_{i,f} + \phi_{i,b})\Omega_i, \qquad (11)$$

which further leads to:

$$\phi_{i,f} + \phi_{i,b} = \phi_{i,0}. \qquad (12)$$

The partition relation described in Eq. (7) can then be written as:

$$(\lambda_{i,b} + \lambda_{i,f})\Delta\phi_{i,b} - nk_B T \ln\frac{\phi_{i,b}}{\phi_{i,0}-\phi_{i,b}} + \frac{1}{2}nk_e|\mathbf{u}|^2 + n(\mu^0_{i,f} - \mu^0_{i,b}) = 0. \qquad (13)$$

By assuming the model parameters are insensitive to different individual cells, the above equation can be written as:

$$(\lambda_b+\lambda_f)\Delta\phi_b - nk_BT\ln\frac{\phi_b}{\phi_0-\phi_b} + \frac{1}{2}nk_e|\mathbf{u}|^2 + n(\mu_f^0 - \mu_b^0) = 0. \qquad (14)$$

We shall note here that once we undistinguish the model parameters between different individual cells, the variable $\phi_{i,b}$ automatically lumps over the whole epithelial sheet.

## 3. The choice of the model parameters

Before we approach, we would like to briefly discuss on how we choose the model parameters. For the parameters used in the partition of the receptors, the gradient energy coefficient $\lambda_f$ and $\lambda_b$ are both chosen to be $1.6 \times 10^{-8}$ $k_BT \cdot m^2$ to ensure a smooth decay in the spatial distribution of $\phi_b$ across the epithelial sheet. The number of the lattice sites per unit area on cell membrane is estimated to be $4 \times 10^4$ μm$^{-2}$, which corresponds to integrins whose cross-section diameter is around 5 nm.(26) The difference in the referential chemical potential between the unbound phase and the bound phase is set to be $\Delta\mu = 5$ $k_BT$ so that $\phi_f/\phi_b \approx 150$ in the reference configuration.(27) The effective spring constant $k_e$ is estimated in the range of $10^{-1}\sim10^2$ $k_BT \cdot$μm$^{-2}$ to ensure that the elastic stretch energy in one ligand-integrin complex in in the range of $0\sim10^2$ $k_BT$ as reported in previous literatures(28). The normalized total density of integrins $\phi_0$ should be much smaller than 1, and is chosen to be 0.1. The parameters used for the mechanical equilibrium of the cell colony are set based on previous works: the Young's modulus $E$ of the cell colonies is usually has the order of magnitude $10^3\sim10^4$Pa;(29,30) the Poisson's ratio $\nu$ often ranges from $0.3\sim0.5$;(16,17,23) the thickness of a monolayer $h_c$ usually ranges from a few microns to approximate ten microns, and the value of the active stress $\sigma_A$ is roughly $10^2\sim10^3$ Pa.(16,25) Therefore, the parameters used here for the mechanical equilibrium are set to be: $E = 10^4$ Pa, $\nu = 0.4$, $h_c = 10$μm, $\sigma_A = 500$ Pa.

## 4. The approximate solution to the weak epithelial-substrate coupling

In this section, we solve the simplified model with respect to the weak epithelial-substrate coupling case where $k_e$ is small. The partition relation in Eq. (14) then becomes:

$$\lambda\left(\frac{d^2}{dr^2}+\frac{1}{r}\frac{d}{dr}\right)\phi_b - \ln\frac{\phi_b}{\phi_0-\phi_b} + \frac{k_e}{2k_BT}(u_r)^2 - \Delta\mu = 0, \tag{15}$$

where $\lambda = (\lambda_b+\lambda_f)/nk_BT$, and $\Delta\mu = \mu_b^0 - \mu_f^0$. The corresponding boundary condition for $\phi_b$ is:

$$\nabla\phi_b \cdot \mathbf{n} = 0 \text{ at } r = 0 \text{ \& } R. \tag{16}$$

The mechanical equilibrium equation written in the radial displacement $u_r$ is:

$$\frac{d^2u_r}{dr^2}+\frac{1}{r}\frac{du_r}{dr} - \left[\frac{nk_e\phi_b(1-\nu^2)}{Eh_c}+\frac{1}{r^2}\right]u_r = 0, \tag{17}$$

whose corresponding boundary conditions are:

$$u_r = 0 \text{ at } r = 0, \tag{18}$$

$$\frac{E}{1-\nu^2}\left[\frac{du_r}{dr}+\nu\frac{u_r}{r}\right] = -\sigma_A \text{ at } r = R. \tag{19}$$

To study how cohesive epithelial monolayer colonies behave on substrate, we solve Eq. (15) and (17) along with their boundary conditions Eq. (16), (18), and (19) semi-analytically by the perturbation method. We expand $\phi_b$ and $u_r$ in the following way:

$$\phi_b = \phi_b^0 + \frac{k_e}{Eh_c}\phi_b^1 + \left(\frac{k_e}{Eh_c}\right)^2\phi_b^2 + o\left[\left(\frac{k_e}{Eh_c}\right)^2\right], \tag{20}$$

$$u_r = u_r^0 + \frac{k_e}{Eh_c}u_r^1 + \left(\frac{k_e}{Eh_c}\right)^2 u_r^2 + o\left[\left(\frac{k_e}{Eh_c}\right)^2\right], \tag{21}$$

where $u_r^0$ satisfies the nonhomogeneous boundary condition Eq. (19), and others satisfy the homogeneous boundary conditions. By substituting Eq. (20) and (21) back into Eq. (15) and (17), respectively, and collecting terms with equal powers of $\frac{k_e}{Eh_c}$, one can obtain a series of equations we need to solve for $\phi_b$ and $u_r$. For the zeroth-order terms on $\frac{k_e}{Eh_c}$, we have:

$$\ln\frac{\phi_b^0}{\phi_0-\phi_b^0} - \Delta\mu = 0, \tag{22}$$

$$\frac{d^2u_r^0}{dr^2}+\frac{1}{r}\frac{du_r^0}{dr}-\frac{u_r^0}{r^2} = 0. \tag{23}$$

The above equations ca be readily solved:

$$\phi_b^0 = \frac{\phi_0}{1+e^{\Delta\mu}}, \tag{24}$$

$$u_r^0 = kr, \tag{25}$$

where the slope $k = -\frac{\sigma_A(1-\nu)}{E}$. The zeroth-order solutions show that for the weak monolayer-substrate coupling where the spring constant of the focal adhesion is small, the density of the integrin receptor tends to be a constant over the whole monolayer colony domain and the displacement in the epithelial sheet can be described by a linear function. Next, for the first-order terms on $\frac{k_e}{Eh_c}$, we have:

$$\lambda \left(\frac{d^2}{dr^2} + \frac{1}{r}\frac{d}{dr}\right)\phi_b^1 - \left(\frac{1}{\phi_b^0} - \frac{1}{\phi_b^0 - \phi_0}\right)\phi_b^1 + \frac{1}{2k_BT}(u_r^0)^2 = 0, \tag{26}$$

$$\frac{d^2 u_r^1}{dr^2} + \frac{1}{r}\frac{du_r^1}{dr} - \frac{u_r^1}{r^2} - \frac{n(1-v^2)}{Eh_c}\phi_b^0 u_r^0 = 0, \tag{27}$$

where we have expanded the function $ln\frac{\phi_b}{\phi_0-\phi_b}$ with respect to $\phi_b$ at $\phi_b = \phi_b^0$. The solutions can be derived analytically as:

$$\phi_b^1 = \frac{\sigma_A^2(1-v)^2 \phi_0 e^{\Delta\mu} h_c}{2(1+e^{\Delta\mu})^2 E k_B T}\left[-2R\frac{\sqrt{\lambda\phi_0 e^{\Delta\mu}}}{1+e^{\Delta\mu}} I_1^{-1}\left(\frac{1+e^{\Delta\mu}}{\sqrt{\lambda\phi_0 e^{\Delta\mu}}}R\right) I_0\left(\frac{r\sqrt{\lambda\phi_0 e^{\Delta\mu}}}{1+e^{\Delta\mu}}\right) + r^2 + \frac{4\lambda\phi_0 e^{\Delta\mu}}{(1+e^{\Delta\mu})^2}\right], \tag{28}$$

$$u_r^1 = \frac{n(1-v^2)(1-v)\sigma_A\phi_b^0}{8E}\left[\frac{(3+v)R^2}{1+v}r - r^3\right]. \tag{29}$$

From Eq. (29), one can see that the first-approximation has a linear part with a slope $k_1$ equal to:

$$k_1 = -\frac{\sigma_A(1-v)}{E}\left[1 - \frac{n(1-v^2)\phi_b^0 k_e}{8Eh_c}\right], \tag{30}$$

which leads to $|k_1| < |k|$. For the second-order terms on $\frac{k}{Eh_c}$, one can have:

$$\lambda\left(\frac{d^2}{dr^2} + \frac{1}{r}\frac{d}{dr}\right)\phi_b^2 - \left(\frac{1}{\phi_b^0} - \frac{1}{\phi_b^0-\phi_0}\right)\phi_b^2 - \frac{1}{2}\left[\frac{1}{(\phi_b^0-\phi_0)^2} - \frac{1}{(\phi_b^0)^2}\right]\phi_b^1 + \frac{Eh_c}{k_BT}u_r^0 u_r^1 = 0, \tag{31}$$

$$\frac{d^2 u_r^2}{dr^2} + \frac{1}{r}\frac{du_r^2}{dr} - \frac{u_r^2}{r^2} - \frac{n(1-v^2)}{Eh_c}(\phi_b^0 u_r^1 + \phi_b^1 u_r^0) = 0. \tag{32}$$

The above equation systems cannot readily be solved analytically, but still can be solved numerically. We plot the solutions to $\phi_b^{1,2}$ and $u_r^{1,2}$ in Fig. 2. We can see that both $\phi_b^{1,2}$ and $u_r^{1,2}$ are positive and monotonically increases as $r$ becomes larger. For $\phi_b^{1,2}$, the solution becomes more localized in the monolayer periphery as the order of the solution increases. We also compare the approximate solutions in Eq. (20) and (21) with the purely numerical solutions to Eq. (8) and (14) for a circular-shaped epithelial monolayer colony with its radius equal to 150 μm, plotted in Fig. 3. The results show that for $k_e \leq 0.16 \text{ k}_B\text{T} \cdot \mu\text{m}^{-2}$, the second-order approximation has a good agreement with the pure numerical solutions for the magnitude of the traction $|T_r|$, the radial component of the displacement $|u_r|$, and the focal adhesion density, represented by $\phi_b$. The errors in the displacement field $|u_r|$ is negligible for all the three values of $k_e$. The errors in the focal adhesion density $\phi_b$ and the magnitude of the traction $|T_r|$, on the other hand, are small in the centroid region of the monolayer colony and gradually grow as the coordinate $r$ increases.

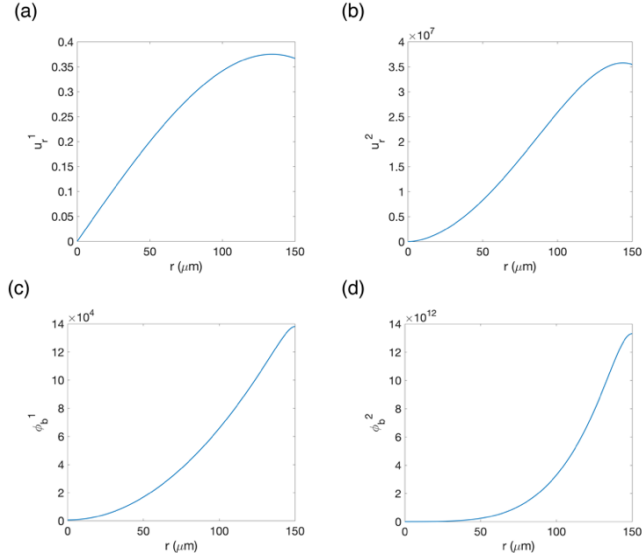

**Fig. 2** The first and second order approximations vs. the radial coordinate: (a) $u_r^1$; (b) $u_r^2$; (c) $\phi_b^1$; (d) $\phi_b^2$.

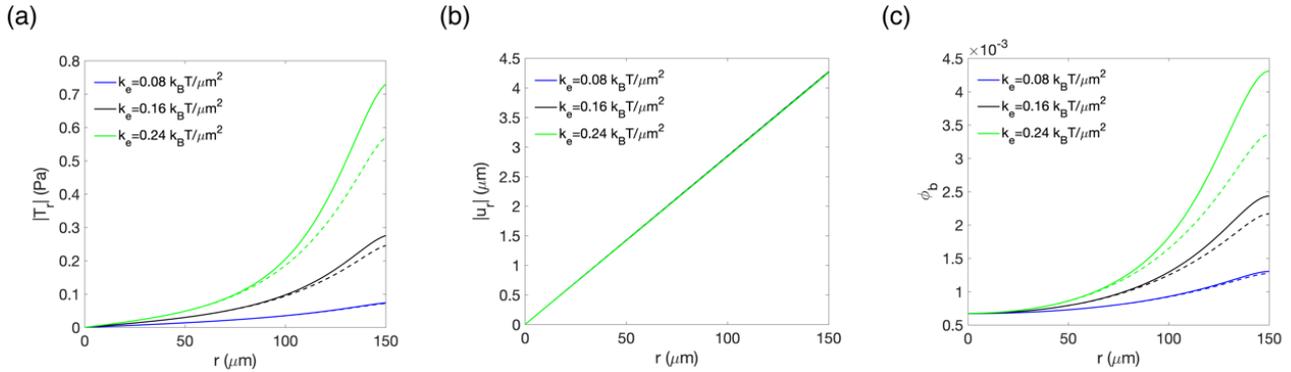

**Fig. 3** Approximated solutions vs. the purely numerical solutions with respect to $k_e = 0.08, 0.16, 0.24 \text{ k}_B\text{T} \cdot \text{μm}^{-2}$ in a circular colony with $R = 150$ μm under the axisymmetric case: (a) the solutions to magnitude of the radial component of traction $|T_r|$; (b) the solutions to magnitude of the radial component of the colony displacement $|u_r|$; (c) the solutions to the normalized density of focal adhesions $\phi_b$. The approximated solutions are sketched in dashed lines, while the numerical solutions are drawn in solid lines. Different colors stand for different values of $k_e$.

## 5. Epithelial-substrate coupling regulates the landscape of the monolayer displacement, focal adhesion distribution, and hence the traction

In this section, we study how epithelial-substrate coupling, characterized by the effective spring constant $k_e$, regulates the spatial distribution of the monolayer displacement **u**, the normalized density of the focal adhesion $\phi_b$, and hence the traction **T**. As a key parameter, $k_e$ can be affected by quite a few factors, especially the stiffness of the substrate: a stiff substrate will lead to a strong coupling between the cells and substrates, while a soft substrate will lead to weak coupling.(10,31) Here, we solve the model in a circular colony with its radius equal to 150 μm with respect to three cases: the weak-coupling case, where $k_e$ is quite small, the strong-coupling case, where $k_e$ is large, and the intermediate cases, where the values of $k_e$ are in between those in the former cases. For the axisymmetric case, we plot the numerical solutions to the radial component of the traction $|T_r|$, the radial component of the displacement $|u_r|$, and the focal adhesion density $\phi_b$ in Fig. 4, 5, and 6, respectively.

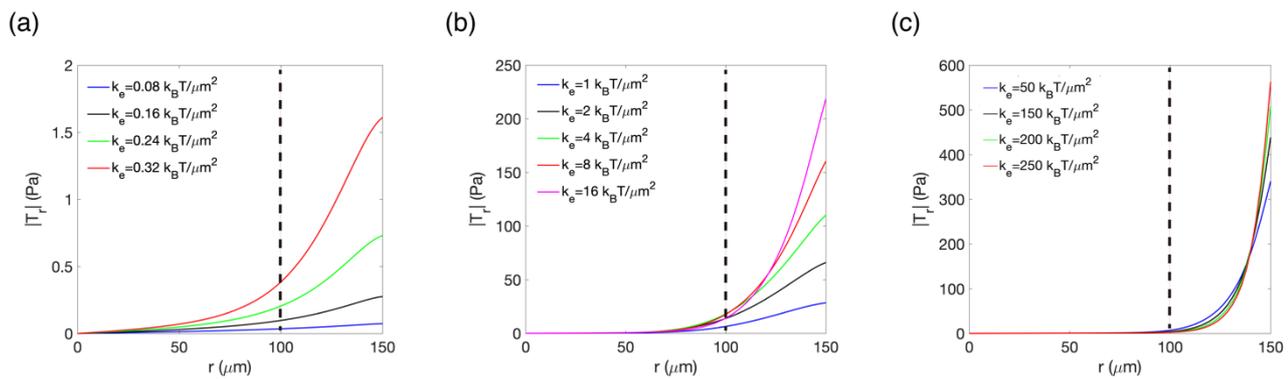

**Fig. 4** The axisymmetric solutions to magnitude of the radial component of traction $|T_r|$ in a circular colony with $R = 150$ μm with respect to cases: (a) the weak epithelial-substrate coupling: $k_e = 0.08$  $0.16, 0.24, \& 0.32$ $k_B T \cdot \mu m^{-2}$; (b) the intermediate epithelial-substrate coupling: $k_e = $

1, 2, 4, 8, & 16 $k_B T \cdot \mu m^{-2}$; (c) the strong epithelial-substrate coupling: $k_e = 50, 150, 200, \& 250\ k_B T \cdot \mu m^{-2}$. The black dashed line indicates a boundary layer with a 50 μm width where the traction are localized for all the three cases.

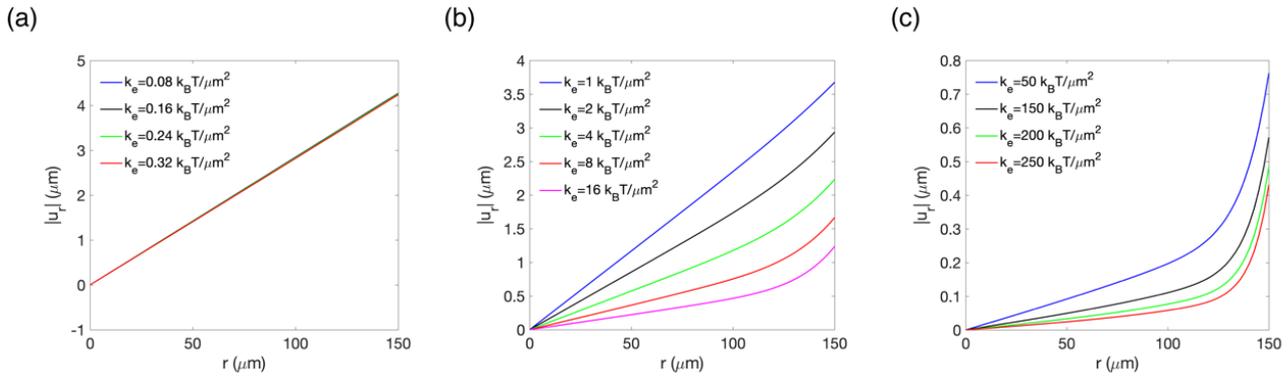

**Fig. 5** The axisymmetric solutions to magnitude of the radial component of displacement $|u_r|$ in a circular colony with $R = 150$ μm with respect to cases: (a) the weak epithelial-substrate coupling: $k_e = 0.08\ 0.16, 0.24, \& 0.32\ k_B T \cdot \mu m^{-2}$; (b) the intermediate epithelial-substrate coupling: $k_e = 1, 2, 4, 8, \& 16\ k_B T \cdot \mu m^{-2}$; (c) the strong epithelial-substrate coupling: $k_e = 50, 150, 200, \& 250\ k_B T \cdot \mu m^{-2}$.

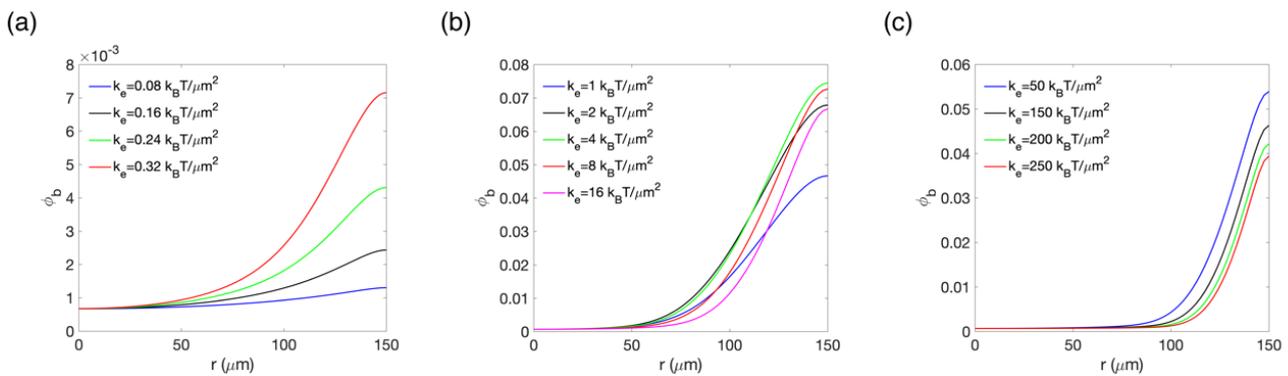

**Fig. 6** The axisymmetric solutions to the normalized density $\phi_b$ of focal adhesions in a circular colony

with $R = 150$ μm with respect to cases: (a) the weak epithelial-substrate coupling: $k_e = 0.08$  $0.16, 0.24,$ & $0.32$ $k_B T \cdot \mu m^{-2}$; (b) the intermediate epithelial-substrate coupling: $k_e = 1, 2, 4, 8,$ & $16$ $k_B T \cdot \mu m^{-2}$; (c) the strong epithelial-substrate coupling: $k_e = 50, 150, 200,$ & $250$ $k_B T \cdot \mu m^{-2}$.

By observing all the three figures, we could see that the values of the traction, the displacement, and the focal adhesion density all reach their maximums at the edge of the epithelial colony, and then approach to zero as the coordinate $r$ arrives at the origin, where the center of the colony is located.(23,32) To be more specific, we firstly look into the profiles of the traction $|T_r|$ with respect to the three cases (Fig. 4). For the weak-coupling and intermediate-coupling cases, the magnitude of the traction at each location inside the monolayer monotonically increases as the value of $k_e$ increases, suggesting that an increasing in the epithelial-substrate coupling strength lead to larger traction in these two cases. For the strong-coupling case, where the values of $k_e$ are much larger than the other two cases, we could see that as the value of $k_e$ increases, the traction tends to be localized more and more in the periphery regions of the monolayer, leading to a growth in $|T_r|$ within the boundary region of the monolayer ($r \geq 140$ μm), yet a decrease for $|T_r|$ in the region just behind it ($100$ μm $\leq r \leq 140$ μm). Such observation indicates that a stronger epithelial-substrate coupling will lead to a faster decay in the magnitude of the traction from the periphery to the centroid region of the monolayer. By observing the results for all the three cases, a boundary layer (shown by the dashed line), with a mutual width of approximately $50$ μm, can be defined based on the localization of traction, which agrees with previous experimental observations(23–25).

Next, we look into the landscapes of the displacement field $|u_r|$ in the monolayer colony (Fig. 5). In the weak-coupling cases, the magnitudes of the displacement $|u_r|$ all behave like linear functions (can be approximated by $u_r^0$) since traction in this case are negligible, resulting in a linear-function solution to the mechanical equilibrium, shown in Eq. (25). As the epithelial-substrate coupling becomes stronger, the magnitude of the displacement field in the periphery region of the colony gradually loses its linearity and becomes more and more curved. The part in the centroid region of the monolayer, on the other hand, still remains linear since there is almost no traction, but the slopes becomes less and less steep. The magnitude of $|u_r|$ also decreases as the epithelial-substrate coupling becomes stronger since the signs of the high-order perturbated terms $u_r^{1,2}$ are opposite from that of $u_r^0$.

At the end of this section, we present the distribution of the focal adhesion with respect different epithelial-substrate coupling strength in Fig. 6. For the case that the value of $k_e$ is smaller than $4\,\mathrm{k_B T} \cdot \mathrm{\mu m^{-2}}$, the density of the focal adhesion gradually increases as $k_e$ increases, implying that cells tend to form more focal adhesions as the coupling strength increases to a certain level. This trend lasts till the value of $k_e$ reaches $4\,\mathrm{k_B T} \cdot \mathrm{\mu m^{-2}}$, beyond which the density of focal adhesion gradually decreases as $k_e$ continues increasing. Such behavior of the focal adhesion density with respect to $k_e$ is then confirmed by Fig. 7, where we plot the average focal adhesion density over the whole epithelial monolayer as a function of the coupling strength $k_e$. The decrease in focal adhesion density happens because the magnitude of the displacement decreases, leading to less energy released from the binding of the integrin to the ligand. As the energy release $\gamma$ becomes smaller, fewer focal adhesion assemblies are in favored.

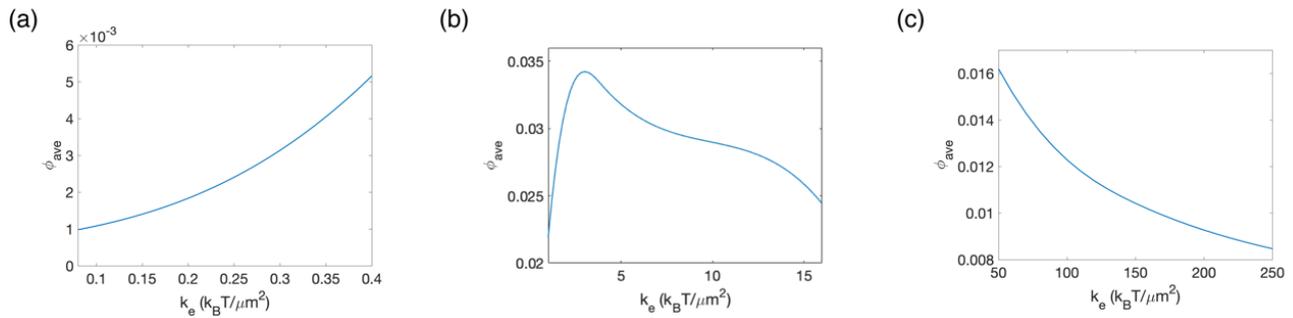

**Fig. 7** The average of the normalized focal adhesion density over the entire epithelial colony as a function of the strength of epithelial-substrate coupling $k_e$ with respect to three cases: (a) the weak-coupling case, where $k_e$ is quite small; (b) the intermediate cases, where the values of $k_e$ are neither too large nor too small; (c) the strong-coupling case, where $k_e$ is very large.

## 6. Size effect: average traction vs. sizes of epithelial monolayers

In this section, we firstly study how the average traction $|T_r|_{ave}$ within the boundary layer changes as a function of the radius of the circular epithelial monolayer, with respect to different strengths of epithelial-substrate coupling, where $|T_r|_{ave}$ is defined as:

$$|T_r|_{ave} = - \int_{R-l}^{R} n\phi_b u_r r \mathrm{d}r \Big/ \int_{R-l}^{R} r \mathrm{d}r, \tag{33}$$

with $l \approx 50\ \mu m$ being the boundary layer width. By looking into Fig. 8 (a), we could see that for an epithelial-substrate coupling strength $k_e \leq 24\ k_B T \cdot \mu m^{-2}$, the average traction within the boundary layer monotonically increases as the radius of the monolayer grows. The growth rate $\Delta|T_r|_{ave}/\Delta R$ gradually decreases as the value of $k_e$ increase. And as the value of $k_e$ becomes larger, the average traction will stop increasing at some certain values of the monolayer radius and start decreasing as the size of the monolayer grows. For $k_e = 28\ \&\ 32\ k_B T \cdot \mu m^{-2}$, such transition happens at $R \approx 130\ \&\ 80\ \mu m$, respectively (Fig. 8 (b)). For the value of $k_e$ that is large enough, the average traction within the boundary layer monotonically decreases as the radius of the monolayer grows (Fig. 8 (c)). The decreasing trend becomes more significant as the coupling strength $k_e$ increases, which finally conforms with the experimental observations[25]. The overall behavior that $|T_r|_{ave}$ vs. $R$ with respect to different values of $k_e$ can be explained in the following way. The average traction $|T_r|_{ave}$ can be approximated as:

$$|T_r|_{ave} = - \int_{R-l}^{R} n \left[ \phi_b^0 + \frac{k_e}{Eh_c}\phi_b^1 + \left(\frac{k_e}{Eh_c}\right)^2 \phi_b^2 \right] \left[ u_r^0 + \frac{k_e}{Eh_c}u_r^1 + \left(\frac{k_e}{Eh_c}\right)^2 u_r^2 \right] r \mathrm{d}r \Big/ \int_{R-l}^{R} r \mathrm{d}r, \tag{34}$$

where the leading term is:

$$|T_r|_{ave}^{lead} = - \int_{R-l}^{R} n(\phi_b^0 u_r^0 + \frac{k_e}{Eh_c}\phi_b^1 u_r^0) r \mathrm{d}r \Big/ \int_{R-l}^{R} r \mathrm{d}r. \tag{35}$$

By substituting Eq. (24) and (25) into Eq. (34), we can find that $|T_r|_{ave}^{lead}$ monotonically increases with $R$. That explains why the average traction within the boundary layer increases as the size of the monolayer increases for small values of $k_e$. As the value of $k_e$ increases, high order terms become more dominant, and it is not difficult to see that $-\int_{R-l}^{R} n\phi_b^{1,2} u_r^{1,2} r \mathrm{d}r \Big/ \int_{R-l}^{R} r \mathrm{d}r$ monotonically decreases as $R$ increases since both $\phi_b^{1,2}$ and $u_r^{1,2}$ have higher order terms on $r$ and they have the same sign. This could explain why the average traction within the boundary layer finally decreases as the size of the monolayer increases for large values of $k_e$. Such conclusion holds for monolayers with other geometries, such as square-shaped epithelial monolayer, plotted in Fig. 9.

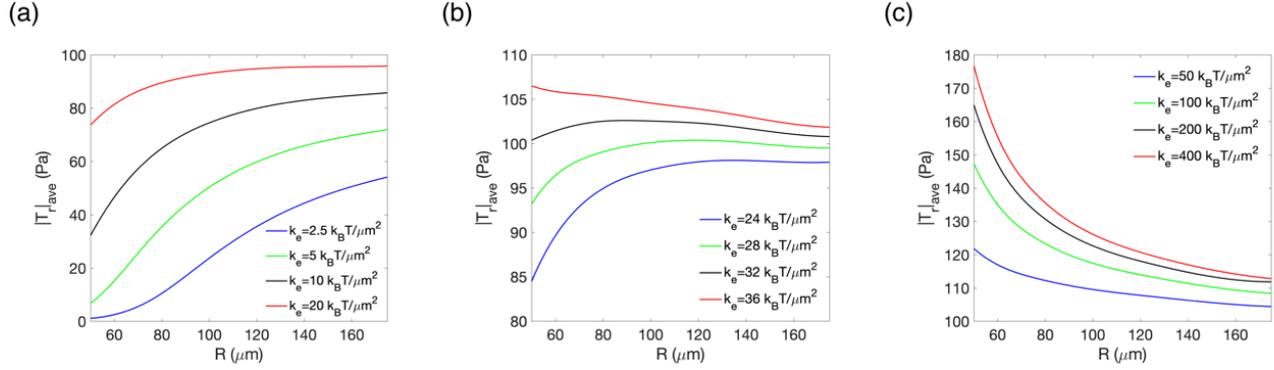

**Fig. 8** The average of the traction $|T_r|_{ave}$ within the boundary layer of circular-shaped colonies as a function of the colony radius with respect to three cases: (a) the weak-coupling case, where $k_e$ is quite small; (b) the intermediate cases, where the values of $k_e$ are neither too large nor too small; (c) the strong-coupling case, where $k_e$ is very large.

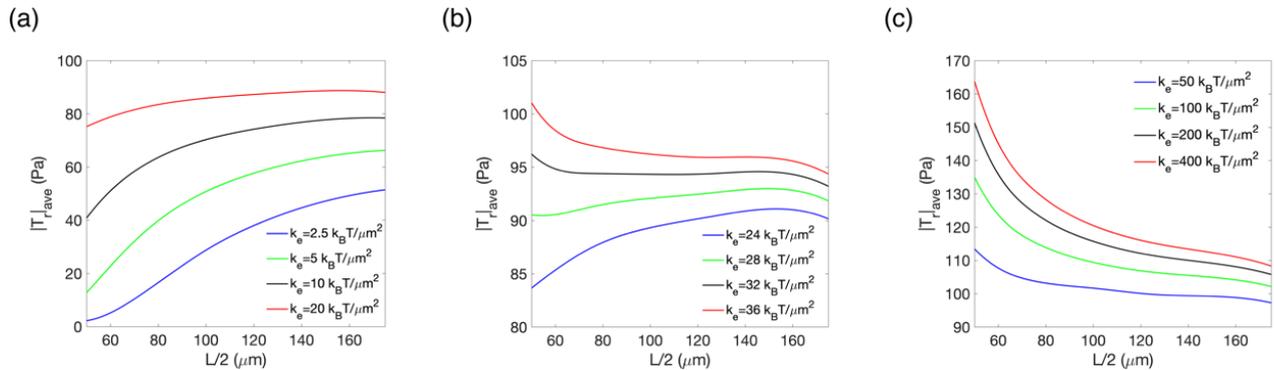

**Fig. 9** The average of the traction $|T_r|_{ave}$ within the boundary layer of square-shaped colonies as a function of the half-edge-length $L/2$ with respect to three cases: (a) the weak-coupling case, where $k_e$

is quite small; (b) the intermediate cases, where the values of $k_e$ are neither too large nor too small; (c) the strong-coupling case, where $k_e$ is very large.

Previous studies have reported a size effect in the nearly-rounded cohesive epithelial monolayers and explained such effect by the possible existence of the large line tension, as mentioned in the introduction part. Based on our computational results in Fig, 8 & 9 (c), we conclude that this experimental observation can be predicted by our model with a strong enough strength of epithelial-substrate coupling ($\sim 10^2$ $k_B T \cdot \mu m^{-2}$ after other parameters being set up); the effect of the line tension, thus, has no need to be introduced. This brings us a question: is the size effect caused by the effect of line tension or is it just a natural outcome by a proper strength of epithelial-substrate coupling? For the line tension effect, previous works have determined that a line tension with a value in the order of $\sim 10^{-7}$ J/m could lead to the size effect observed in the experiment.(25) Yet whether the value of the line tension could reach such value still requires experimental measurements. The boundary line tension $\Gamma$ could arise from two parts: one is the surface tension in the cell lipid membrane; the other is the contractile force in the actin-ring along the monolayer boundary. The former should be a material constant $\gamma_m$, and the latter is proportional to the circumference of the monolayer $\sim fR$, where $f$ is the contractility stiffness in the actin cable.(33,34) The complete inward pressure induced by the line tension then should be $p \sim \frac{\gamma_m}{R} + f/h_c$, where only the membrane term shows the size effect. The surface tension for epithelial cell membrane usually possesses a value in the order of $10^{-4} \sim 10^{-3}$ J/m$^2$,(35) leading to an effective inward pressure which is one-order lower than the value given by the previous works(25). Thus, it is possible that previous works have overestimated the effect brought by the line tension, and further experiments need to be conducted to examine the true mechanism underlying the "size effect".

Finally, we would like to estimate the value of $k_e$ that could lead to the "size effect" observed in the previous experiment. To make it simple for this highly nonlinear equation system, we assume the average

traction obtained from the first order approximation:

$$|T_r|_{ave} = -\int_{R-l}^{R} n\left[\phi_b^0 + \frac{k_e}{Eh_c}\phi_b^1\right]\left[u_r^0 + \frac{k_e}{Eh_c}u_r^1\right]r\mathrm{d}r \Big/ \int_{R-l}^{R} r\mathrm{d}r, \qquad (36)$$

to monotonically decrease as the monolayer radius $R$ increases. To lower down the mathematical complexity, we only keep the lower-order terms with depends on $r$ in $\phi_b^1$ and $u_r^1$:

$$\phi_b^1 \sim \frac{\sigma_A^2(1-\nu)^2\phi_0 e^{\Delta\mu}h_c}{2(1+e^{\Delta\mu})^2 Ek_BT}r^2, \qquad (37)$$

$$u_r^1 \sim \frac{n(1-\nu^2)(1-\nu)\sigma_A\phi_b^0}{8E}\left[\frac{(3+\nu)R^2}{1+\nu} r\right]. \qquad (38)$$

We note here, the higher-order terms in in $\phi_b^1$ and $u_r^1$ only have prominent effect near the monolayer boundary, and neglecting them still preserves the orders of $\phi_b^1$ and $u_r^1$. By substituting Eq. (37) and (38) into Eq. (36), we can find that the critical value of $k_e$ that determine the monotonically decreasing of $|T_r|_{ave}$ with respect to $R$ is:

$$(k_e^{cr})^2 \sim \left[\frac{40Eh_c}{3(3+\nu)(1-\nu)R^4\varepsilon_A^2}\right]\frac{k_BT}{n\phi_b^0}, \qquad (39)$$

where $\varepsilon_A$ is the active strain defined as: $\varepsilon_A = \frac{\sigma_A(1-\nu)}{E}$. By letting $\nu \approx 0.5$, Eq (39) is further simplified to:

$$(k_e^{cr})^2 \sim \frac{8Eh_c}{R^4\varepsilon_A^2}\frac{k_BT}{n\phi_b^0}. \qquad (40)$$

We can see that both the macroscopic properties and the microscopic thermodynamics properties of the cohesive monolayer contribute to the critical value $k_e^{cr}$. By substituting the parameters (with $R_{min} = 50$ μm) used in the paper, we obtained the value of $k_e^{cr} \sim 38\,\mathrm{k_BT\cdot\mu m^{-2}}$, which is a little larger than the value $k_e = 36\,\mathrm{k_BT\cdot\mu m^{-2}}$ shown in Fig. 9(b). Moreover, we can tell from Eq. (40), stiffer and thicker monolayers need stronger coupling with substrates to show the size effect observed by previous experiments. On the other hand, monolayers with large active strains and referential ratio of $\phi_b/\phi_f$ are more likely to be observed the size effect.

7. **Conclusion.**

In this paper, we have described a thermodynamic steady-state model that characterizes the cohesive epithelial-substrate interaction in details. In the model, a free energy functional that incorporates the elastic strain energy in the epithelial monolayer and the chemical energy in the integrins is formulated. The thermodynamic steady-state equations within each cell are derived by the minimization of the functional with respect to its independent variables. The equation system is simplified based on the fact that the individual cell size is much smaller than the entire colony size. We then solve the model with respect to different strength of epithelial-substrate coupling, which is characterized by the parameter $k_e$. We have found that for the weak epithelial-substrate coupling where $k_e$ is small, the model can be solved semi-analytically. The landscapes of the monolayer displacement $\mathbf{u}$, the normalized density of the focal adhesion $\phi_b$, and hence the traction force $\mathbf{T}$ can be quite different with respect to different values of $k_e$. Most importantly, we have found that the strong epithelial-substrate coupling alone can fully explain the size effect of cohesive epithelial colonies observed from previous experiments. The effect of the line tension, which is introduced by previous models to explain such effect, has no need to be considered. This founding holds for different geometries, such as square-shaped colonies. We also have found that the relatively weak epithelial-substrate coupling could lead to an inverse size effect, which requires further experimental examinations. At last, a scaling law of the epithelial-substrate coupling strength that could lead to the "size effect" observed from previous works has been proposed.

**Author Contribution**

T. Z. performed the model derivation and computation; T. Z. performed the discussion; manuscript was prepared by T.Z. and H. Y.

**Conflicts of Interest**

The authors declare no competing financial interest.

**Acknowledgements**

The authors would like to acknowledge the funding support from the Science, Technology and Innovation Commission of Shenzhen Municipality (grant no. ZDSYS20210623092005017).